# Fe-carbon nitride *"Core-shell"* electrocatalysts for the oxygen reduction reaction

Keti Vezzù[1], Antoine Bach Delpeuch[1], Enrico Negro[1,2,3], Stefano Polizzi[4], Graeme Nawn[1], Federico Bertasi[1,3], Gioele Pagot[1,2], Kateryna Artyushkova[5], Plamen Atanassov[5], Vito Di Noto[1,3,6*]

[1] *Section of "Chemistry for the Technology" (ChemTec), Department of Industrial Engineering, University of Padova, in the Department of Chemical Sciences, Via Marzolo 1, I-35131 Padova (PD), Italy.*

[2] *Centro Studi di Economia e Tecnica dell'Energia «Giorgio Levi Cases», 35131 Padova (PD) Italy.*

[3] *Consorzio Interuniversitario per la Scienza e la Tecnologia dei Materiali (INSTM).*

[4] *Department of Molecular Sciences and Nanosystems and Centre for Electron Microscopy "G. Stevenato", University Ca' Foscari Venice, Via Torino 155/B, I-30172 Venezia-Mestre (VE), Italy*

[5] *Department of Chemical and Biological Engineering, and Center for Micro-Engineered Materials University of New Mexico, Albuquerque, NM, USA*

[6] *CNR-ICMATE, Via Marzolo 1, I-35131 Padova (PD), Italy..*

**e-mail:* vito.dinoto@unipd.it

*Keywords:* Fe-carbon nitride based electrocatalysts; Fe-N-C Platinum Group Metal-free catalysts, Oxygen reduction reaction; *"Core-shell"* morphology; Powder X-ray diffraction; CV-TF-RRDE method.

**Abstract**

In this report, the preparation of Fe-carbon nitride (CN)-based electrocatalysts (ECs) with a *"core-shell"* morphology for the oxygen reduction reaction (ORR) is described. The ECs consist of spherical XC-72R carbon nanoparticles, the *"cores"*, that are covered by a CN matrix, the *"shell"*, embedding Fe species in *"coordination nests"*. The latter is formed by the presence of carbon and nitrogen ligands on the surface of the CN matrix, the *"shell"*. Two families of CN-based ECs are prepared, which are grouped on the basis of the concentration of N atoms in the CN *"shell"*. Each group comprises of both a *"pristine"* and an *"activated"* EC; the latter is obtained from the *"pristine"* EC by a suitable series of treatments (**A**) devised to improve the ORR performance. The chemical composition of the CN-based ECs is determined by Inductively-Coupled Plasma Atomic Emission Spectroscopy (ICP-AES) and microanalysis. The thermal stability under both inert and oxidizing atmospheres is gauged by High-Resolution Thermogravimetric Analysis (HR-TGA). The structure is probed by powder X-ray diffraction, and the morphology is inspected by Scanning Electron Microscopy (SEM) and High-Resolution Transmission Electron Microscopy (HR-TEM). The surface area of the CN-based ECs is determined by nitrogen physisorption techniques, and the surface composition is probed by X-ray Photoelectron Spectroscopy (XPS). The electrochemical performance and reaction mechanism of the CN-based ECs in the ORR is investigated in both acid and alkaline environments by cyclic voltammetry with the Thin-Film Rotating Ring-Disk Electrode setup (CV-TF-RRDE). The influence of the preparation parameters and of the treatments on the physicochemical properties, the ORR performance, and reaction mechanism is studied in detail. In the alkaline environment the $FeFe_2$-$CN_l$ 900/$C_A$ *"core-shell"* EC shows a remarkable ORR onset potential of 0.908 V *vs.* RHE which, with respect to the value of 0.946 V *vs.* RHE of the Pt/C ref., classifies the proposed materials as very promising *"Platinum Group Metal-free"* ECs for the ORR.

## 1. Introduction

As of today, one of the most important global challenges is the necessity to renovate completely the infrastructure required to harness and distribute energy [1-3]. New environment-friendly technologies must be implemented on a large scale to curtail the emissions of greenhouse gases and reduce the dependency on non-renewable energy sources [1]. To achieve these goals, electrochemical energy conversion and storage systems are expected to play a continually increasing role owing to their high efficiency and flexibility [3, 4]. In particular, low-temperature fuel cells (FCs) and metal-air batteries have attracted considerable interest due to their high energy and power densities [5-7], thus making them suitable for small stationary systems [8], the automotive sector [9-11] and portable electronics [12, 13].

The operation of both metal-air batteries and low-temperature FCs is bottlenecked by the sluggish kinetics of the oxygen reduction reaction (ORR) [14, 15]. In the case of low-temperature fuel cells, until very recently, the only viable electrolytes were proton-conducting systems (*e.g.*, perfluorinated ionomers such as Nafion®, Hyflon-Ion®, 3M polymer, hydrocarbon-based membranes, among many others) [16], giving rise to a highly acidic environment at the electrodes. Accordingly, low-temperature FCs had to rely on ORR electrocatalysts (ECs) containing a high loading of platinum-group metals (PGMs) in order to achieve a performance and durability level compatible with the applications [17, 18]. The latter is one of the main reasons why the large-scale rollout of low-temperature fuel cells (*e.g.*, proton-exchange membrane fuel cells, PEMFCs, and direct methanol fuel cells, DMFCs) has not been realized. In recent years, significant progress has been achieved in the development of advanced anion-exchange membranes (AEMs) allowing the facile diffusion of $OH^-$ anions [19, 20]. These systems generate a strongly alkaline environment at the electrodes, which is able to provide a promising ORR kinetics by means of *"Pt-free"* ECs [21-23]. The latter are mostly based on common transition metals, thus lowering the risk of supply bottlenecks [24]. These benefits have prompted significant research efforts, aimed at developing high-performance, *"Pt-free"* ECs for the alkaline environment [25, 26].

There is a multitude of synthetic approaches to obtain *"Pt-free"* ORR electrocatalysts for the alkaline environment. A broad family of ECs includes the pyrolysis of precursors obtained by: (a) the chemical modification of carbon nanostructures (*e.g.*, carbon nanoparticles [27], carbon nanotubes [28, 29], graphene sheets [30, 31]); and (b) natural materials (*e.g.*, blood biomass [29], or eggs [32]). Other ECs are prepared by supporting on carbon black active sites based on inorganic compounds such as oxides [33], sulfides [34], selenides [35] and many others. It should be pointed out that in the last decade, a unique and highly flexible synthetic protocol was devised, and later improved, in our laboratory [36-38], which allowed for the preparation of two generations of metal alloy carbon nitride ECs. The first generation includes ECs based on small metal alloy nanoparticles (NPs) embedded in carbon nitride (CN) matrices of larger NPs [39, 40]. The CN matrix consists of graphitic materials in which a small amount of carbon atoms are substituted by N. The second generation groups the *"core-shell"* morphology ECs (CsECs). CsECs consist of a matrix of CN material, the *"shell"*, which is wrapping homogeneously conductive NPs, the *"core"* [38, 41, 42]. Crucial in the preparation of these two generations of ECs is the synthesis of precursors, which in both cases are typically based on a suitable Zeolitic Inorganic-Organic Polymer Electrolyte (Z-IOPE). The latter is synthesized as reported elsewhere through a sol-gel and gel-plastic transition [37, 39]. The Z-IOPE precursor consists of a 3D hybrid inorganic-organic crosslinked system, where clusters of the desired metals are bonded together through bridges of a suitable binder[38]. The multi-step pyrolysis process of the Z-IOPE precursors plays a crucial role in the modulation of the chemical composition and of the morphology of the active sites of the CN-based ECs[43]. It was proven that the CN-based ECs show an outstanding ORR performance in an acid environment, both *"ex-situ"* and at the cathode of a single PEMFC [37, 44, 45]. The active sites are typically located on the surface of alloy NPs coordinated by the nitrogen and carbon ligands of the *"coordination nests"* of the CN matrix [42, 44, 46]. In the CN-based ECs, the alloy NPs bearing the active sites are stabilized in highly conducting CN matrices where the concentration of N atoms is lower than 5 wt%. Most of these N atoms are located in the *"coordination nests"* of the alloy NPs[37, 38]. In this way, CN-based ECs with a high dispersion of the active sites are easily obtained, which address the issues associated with the mass transport of reactants and products towards and from the active sites of the ECs. In ECs of the proposed composition, a good electrical contact between active sites and the external circuit is guaranteed by the CN matrix [37, 38, 44].

This report for the first time describes the preparation protocol devised and optimized to synthesize ORR Fe-CN-based ECs: (i) demonstrating the feasibility of the proposed preparation protocol to obtain highly active *"Pt-free"* ORR electrocatalysts for the alkaline environment; and (ii) elucidating the complex interplay existing between the preparation parameters, the activation process (**A**), the physicochemical properties and the electrochemical performance of the resulting *"core-shell"* Fe-CN-based ECs.

## 2. Experimental

### 2.1. Reagents

Potassium hexacyanoferrate (II) trihydrate (98%) and iron (III) chloride (97%) are both purchased from Sigma-Aldrich. Sucrose, molecular biology grade is Alfa Aesar reagent. Perchloric acid (67-72%), hydrofluoric acid (48 wt%) and potassium hydroxide (98.4%) are provided by Fluka Analytical, Sigma-Aldrich and VWR International. Isopropanol (> 99.8%) is received from Sigma-Aldrich. All chemicals are used as received, without any further purification procedure. Doubly distilled water is used in all the experiments. EC-10 electrocatalyst is procured from ElectroChem, Inc. (nominal Pt loading: 10 wt%); it is labeled *"Pt/C ref."* throughout the text. XC-72R is supplied by Carbocrom S.r.l. and is washed with $H_2O_2$ (10 vol.%) prior to use.

### 2.2. Synthesis of the electrocatalysts

The synthesis of the new Fe-CN-based ECs, which is inspired by the preparation protocol previously reported [47], is discussed in the following text. $FeCl_3$ (487 mg) is dissolved in a minimum amount of water (~ 2 mL). To this sample is added: (i) in a first step, a viscous solution prepared mixing 1820 mg of sucrose to *ca.* 2 mL of water; and (ii) in a second step, 1820 mg of XC-72R nanoparticles. The resulting suspension (Suspension I) is finally homogenized by probe sonication. A second suspension (Suspension II) is prepared as above described, using 2534 mg of $K_4Fe(CN)_6 \cdot 3H_2O$ in place of $FeCl_3$. Suspension I and Suspension II are mixed, thoroughly homogenized by probe sonication, and allowed to rest for 2 days. The resulting product is dried in an oven at 120°C and transferred into a quartz tube, where it undergoes a three-step pyrolysis process under dynamic vacuum as follows: Step 1: 150°C, 7 hours; Step 2: 300°C, 2 hours; Step III: 900°C, 2 hours. The resulting product is divided into two aliquots, labeled A and B. A is washed three times with water, yielding the *"pristine"* CN-based EC. B undergoes the *"activation process"* (indicated as **A** in the following text), which consists of: (a) an etching step with 10 wt% hydrofluoric acid (HF) lasting two hours, followed by an extensive washing with water; and (b) a pyrolysis step under dynamic vacuum at 900°C lasting two hours. The resulting product is the *"activated"* CN-based EC. The protocol above is used to obtain two groups of CN-based ECs. In the first group, the molar ratio between $FeCl_3$ and $K_4Fe(CN)_6{\cdot}3H_2O$ is 1:2; accordingly, the resulting *"pristine"* and *"activated"* CN-based ECs are labeled *"$FeFe_2$-$CN_l$ 900/C"* and *"$FeFe_2$-$CN_l$ 900/$C_A$"*, respectively. In the second group, the molar ratio between $FeCl_3$ and $K_4Fe(CN)_6{\cdot}3H_2O$ is 2:1; and the synthesis is carried out exactly as above, with the difference that the amounts of $FeCl_3$ and $K_4Fe(CN)_6{\cdot}3H_2O$ are 974 and 1267 mg, respectively. In this latter group, the *"pristine"* and *"activated"* CN-based ECs are labeled *"$Fe_2Fe$-$CN_l$ 900/C"* and *"$Fe_2Fe$-$CN_l$ 900/$C_A$"*, respectively.

**2.3. Instruments and methods**

The assay of C, H, N and S is determined by elemental analysis using an FISONS EA-1108 CHNS-O instrument. The wt% of Fe and K is assessed by Inductively-Coupled Plasma Atomic Emission Spectroscopy (ICP-AES) carried out by means of a SPECTRO Arcos spectrometer with EndOnPlasma torch. The digestion of the samples is described elsewhere [40]. The emission lines are: λ (Fe) = 259.940 nm, λ (K) = 766.490 nm. High-Resolution Thermogravimetric Analyses (HR-TGA) are conducted in the temperature range between 30 and 1000°C using a TGA 2950 analyzer (TA instruments). The sensitivity of the instrument ranges from 0.1 to $2\%{\cdot}min^{-1}$; the resolution is 1 μg. The heating rate is varied on the basis of the first derivative of the weight loss from 50 to $0.001°C{\cdot}min^{-1}$. The measurements are performed with an open Pt pan, both in an inert ($N_2$) and in an oxidizing (air) atmosphere. Powder X-ray profiles are collected with a GNR diffractometer (mod. eXplorer) mounting a monochromatized $CuK_\alpha$ source in the 2θ range 3-70° with a 0.05° step and integration time of 40 sec. The XRD profiles are analyzed with the MAUD software [48]. A Cambridge Stereoscan 250 Mark 1 scanning electron microscope, operating at an acceleration voltage of 20 kV, is used to collect standard images with backscattered electrons. Both conventional and high-resolution transmission electron microscopy studies are carried out with a Jeol 3010 apparatus equipped with a high-resolution pole piece (0.17 nm point-to-point resolution) and a Gatan slow-scan 794 CDD camera. The preparation of the samples takes place in accordance with a protocol described elsewhere [44]. XPS spectra are collected at a Kratos Axis DLD Ultra X-ray photoelectron spectrometer using an Al $K_\alpha$ monochromatic source operating at 150 W with no charge compensation. The base pressure is *ca.* $2\cdot10^{-10}$ Torr; the operating pressure is *ca.* $2\cdot10^{-9}$ Torr. Survey and high-resolution spectra are acquired at pass energies of 80 and 20 eV, respectively. Acquisition time for survey spectra is 3 min; in the case of high-resolution spectra, the acquisition time is 2 min for C1s, 5 min for O1s, and 30 min for both N1s and Fe2p regions. The analysis and quantification of the data are carried out in CasaXPS. The C1s, O1s and N1s spectra are quantified using linear background while Fe 2p using Shirley background subtraction. Sensitivity factors provided by the manufacturer are utilized. The curve-fit of N1s spectra is done using a 70% Gaussian/30% Lorentzian line shape. The surface area of the samples is determined by a Micrometrics Gemini V system using a 5-points Brunauer–Emmett–Teller (BET) method.

### 2.4. Electrochemical measurements

The electrode inks are prepared following a protocol described in the literature [39]. Each electrode ink is deposited onto the glassy carbon (GC) disk (Ø 0.5 mm) of the rotating ring-disk electrode (RRDE) tip. The total loading of CN-based EC on the GC disk of the RRDE tip is equal to 0.765 $mg\cdot cm^{-2}$. In the case of the Pt/C ref., the Pt loading is 15 $\mu g_{Pt}\cdot cm^{-2}$. The electrochemical setup is described elsewhere [41]. Measurements in acid and alkaline environments are collected using as electrolyte a 0.1 M $HClO_4$ and a 0.1 M KOH solution, respectively. The $Hg/HgSO_4/K_2SO_{4(sat.)}$ and $Hg/HgO/KOH_{(aq)}$(0.1 M) reference electrode for the acid and the alkaline environment, respectively, is placed in a separate compartment. All potentials are reported in terms of the reversible hydrogen electrode (RHE) scale. Calibration is carried out by $H_2$ oxidation/reduction measurements on a platinized platinum electrode before each experiment. The rotating ring-disk working electrode is mounted on a Model 636 rotator (Pine Research Instrumentations). The collection efficiency of the Pt ring is equal to 0.38. The data are collected with a multi-channel VSP potentiostat/galvanostat from Bio-Logic. CN-based ECs are activated electrochemically by cycling the CN electrode films between $E$ = 0.05 and 1.05 V *vs.* RHE in an $O_2$-saturated 0.1 M $HClO_4$ solution at a sweep rate ($v$) of 100 $mV \cdot s^{-1}$ and with a tip rotation of the RRDE electrode of 1600 rpm until the voltammogram is stabilized. The electrochemical measurements are then collected at the same RRDE rotation rate and at $v = 20$ $mV \cdot s^{-1}$. During the measurements, the ring electrode is always maintained at $E$ = 1.2 V *vs.* RHE to detect the hydrogen peroxide [49]. The faradic ORR currents are obtained measuring first the cyclic voltammogramms in the $N_2$-saturated 0.1 $HClO_4$, and then subtracting these values from those collected under an oxygen atmosphere [50]. The contribution of the ohmic resistance of the electrochemical setup, which is evaluated after each measurement, is removed from the cyclic voltammograms [51]. To evaluate the activity of the CN-based ECs in different atmospheres, high purity nitrogen and oxygen gases (Air Liquide) are used to saturate the electrochemical cell. Tests in alkaline conditions are performed using the same procedure but with a 0.1 M KOH electrolyte solution. Current densities are obtained normalizing the disk currents to the GC geometric area.

## 3. Results and Discussion

The ORR performance of the proposed ECs is maximized by: (i) adopting a *"core-shell"* morphology with *"cores"* of spherical XC-72R carbon black NPs and *"shells"* of CN matrix; (ii) introducing a low concentration of N in the CN matrix, to be mostly included in *"coordination nests"*; and (iii) taking advantage of Fe-CN-based coordination species to promote the electrochemical process. Two groups of CN-based ECs are prepared, distinguished by the concentration of cyano ligands in the starting metal complexes. Each group of CN-based ECs comprises both a *"pristine"* and an *"activated"* material. Fe is studied owing to its ability to provide stable metal-carbon coordination bonds with the CN matrix and for its well-known capability to promote the ORR both in the acid and in the alkaline medium [52]. With the aim to improve the ORR performance, *"activated"* ECs (which include the subscript *"A"* in the labels) are prepared by suitable treatments of *"pristine"* ECs to remove *"non-active"* Fe-based species and contaminants.

### 3.1. Chemical Composition

TABLE 1

All the CN-based ECs comprise a small wt% of heteroatoms, *i.e.* hydrogen, nitrogen, and sulfur. The chemical analysis of the CN-based ECs is summarized in Table 1. The presence of H suggests that the graphitization of the organic fraction of the precursor is incomplete despite the high temperature of the final step of the pyrolysis process ($T_f$ = 900°C) [40]. N is introduced in the CN *"shell"* matrix exclusively by the cyano ligands of the $K_4Fe(CN)_6{\cdot}3H_2O$ reagent complexes. On these bases, it is reasonable to assume that N is not homogeneously distributed throughout the *"shell"* matrix, but it is mostly located close the Fe species [36], which are thus stabilized in *"coordination nests"* of CN matrix based on C- and N-ligands [44]. In accordance with the stoichiometry of the reagents, the N fraction in $FeFe_2$-$CN_l$ 900/C is almost twice that of $Fe_2Fe$-$CN_l$ 900/C. Traces of sulfur are detected which, as expected, are introduced by the XC-72R support [53]. After the extensive pyrolysis and washing processes of ECs, potassium is still present in the materials. This evidence demonstrates that negatively-charged Fe coordination complexes and alcoholate/carboxylate functional groups [42] are present in the CN *"shell"*, which are neutralized by $K^+$ cations [40]. The presence of these ionic groups likely improves the hydrophilicity of the resulting ECs.

The concentration of Fe in both pristine ECs is very similar and is in accordance with the starting stoichiometry of the reagents, while **A** process significantly influences the stoichiometry of the ECs as follows: (a) the wt% of H is reduced, witnessing an improvement in the graphitization of the *"activated"* ECs; (b) the wt% of Fe rises in the order $Fe_2Fe$-$CN_l$ 900/$C_A$ < $FeFe_2$-$CN_l$ 900/$C_A$, in accordance with the trend of the wt% of N in the corresponding pristine ECs (see Table 1). This result highlights that the N-based ligands of the *"coordination nests"* of CN matrix in pristine ECs act to stabilize the metal complexes (see Section 3.3 and Section 3.4). Indeed, after **A** process the Fe species less stabilized owing to coordination by N ligands in the *"coordination nests"* are easily removed by the etching with HF. Thus, the type and density of the N-ligand species forming the *"coordination nests"* of CN matrices play a crucial role in the chemical stability of Fe NPs and of the Fe-coordination complexes.

### 3.2. HR-TG analyses

The HR-TGA profiles reported in Figure 1(a) and Figure 1(b) allow us to study the thermal stability of the CN-based ECs, Pt/C ref. and XC-72R carbon black in inert ($N_2$) and oxidizing (air) atmospheres.

FIGURE 1

It should be noticed that under an $N_2$ atmosphere the thermal stability at T > 600°C (see Figure 1(a)) of the CN-based ECs is higher than that of the Pt/C ref. This evidence, which is in accordance with other studies [37], demonstrates that the presence of nitrogen in the CN matrix *"shell"* improves the thermal stability of ECs in the order: $Fe_2Fe$-$CN_l$ 900/C < $FeFe_2$-$CN_l$ 900/C (see Table 1). Two thermal events (I and II) are observed in the HR-TGA profiles of ECs (see Figure 1(a)). Between *ca.* 550 and 700°C, (I) is attributed to the decomposition of the CN *"shell"*; and (II), at T > 800°C, is assigned to the degradation of the XC-72R *"cores"*. (II) is promoted by the presence of metal species on the surface of the ECs. Moreover, **A** improves the thermal stability of ECs increasing the graphitization of the CN *"shell"* [37, 44]. The mass loss at T < 200°C is diagnostic of the hydrophilicity of the ECs, and corresponds to the elimination process of $H_2O$. As expected, the hydrophilicity of the ECs rises on the N concentration and decreases on **A**, as follows: XC-72R < Pt/C ref. < $Fe_2Fe$-$CN_l$ 900/$C_A$ < $FeFe_2$-$CN_l$ 900/$C_A$ < $Fe_2Fe$-$CN_l$ 900/C < $FeFe_2$-$CN_l$ 900/C.

In the oxidizing atmosphere, the main degradation event $T_D$ is detected between 400 and 680°C. $T_D$ is associated to the oxidation of the carbon-based matrix (see Figure 1(b)), and rises in the order: $T_D$ (Pt/C ref.) ≈ $T_D$ ($FeFe_2$-$CN_l$ 900/C) < $T_D$ ($Fe_2Fe$-$CN_l$ 900/C) < $T_D$ (XC-72R). This behavior is easily explained if we consider that the $O_2$ of air preferentially adsorbs on Fe complexes stabilized by N-ligands of the CN matrix, thus facilitating the combustion of the whole EC. Indeed, a higher density of N-ligands in *"coordination nests"* improves the oxophilicity and the concentration of the Fe-coordination species of the CN matrix. This confirms the result that the N concentration in ECs decreases in the order: $FeFe_2$-$CN_l$ 900/C (N wt% = 0.74, $T_D$ ≈ 400°C) > $Fe_2Fe$-$CN_l$ 900/C (N wt% = 0.38, $T_D$ ≈ 550°C) > XC-72R (N wt% = 0, $T_D$ ≈ 680°C). The relatively low $T_D$ value of the Pt/C ref. demonstrates that Pt NPs are efficient catalysts for the degradation of the carbon support[54]. The correlation existing between $T_D$ and **A** process (see Figure 1(b)) highlights that the CN-based ECs include two main types of Fe species: (i) oxophilic Fe species which, thanks to their stabilization in *"coordination nests"* (see above), are not affected significantly by **A** and modulate $T_D$; and (ii) other

*"inert"* Fe species, not stabilized in *"coordination nests"*, which are easily removed during **A** and do not affect $T_D$. The HR-TGA profile in the high-temperature wing corresponds to the $Fe_xO_y$ residue under an oxidizing atmosphere. The wt% of this latter residue is in accordance with the Fe assay (see Table 1) and in ECs rises in the order: $FeFe_2$-$CN_l$ 900/C ≈ $Fe_2Fe$-$CN_l$ 900/C > $FeFe_2$-$CN_l$ 900/$C_A$ > $Fe_2Fe$-$CN_l$ 900/$C_A$.

### 3.3. Powder X-Ray Diffraction Studies

The powder X-ray diffraction (XRD) profiles of the ECs shown in Figure 2 are quantitatively analyzed by the Rietveld method in order to identify as much as possible the structural features of the different phases composing the ECs and to obtain an estimate of their domain size and relative abundance [55]. For the sake of comparison the XRD profile of XC-72R, together with the results of its Rietveld analysis, are shown in Figure S1 of Supplementary Information.

FIGURE 2

TABLE 2

The phase characterizing the XRD patterns of the CN-based ECs consists of a hexagonal $P6_3/mc$ component corresponding to the XC-72R *"core"* support with domain sizes in the range of *ca.* 2.2-2.5 nm, COD#9008569 [56]. In details, the hexagonal $P6_3/mc$ phase of XC-72R is characterized by a main (002) peak at 2θ ≈ 25° and by two minor peaks at 2θ ≈ 43°, 56° which are associated with the (111) and (004) reflections, respectively. A careful analysis of the XRD profiles of ECs demonstrates that the preparation procedure has no influence on the structure and the morphology of this phase. Peaks ascribed to Fe-based NPs are detected, which are quite sharp and correspond to metal NPs with sizes larger than 80-90 nm. These Fe-based NPs, which are coordinated by the CN *"shell"* matrix, are formed during the pyrolysis process from clusters of Fe coordination complexes of the precursor, in accordance with the mechanism elsewhere described[38]. The metal component mostly consists of the following species: (a) in $FeFe_2$-$CN_l$ 900/C: cementite ($Fe_3C$, COD#9012188 [56]) for *ca.* 10 wt%; (b) in $Fe_2Fe$-$CN_l$ 900/C: magnetite ($Fe_3O_4$, COD#1010369 [56]) for *ca.* 12 wt%, with some traces of cementite. These results prove that nitrogen concentration in the CN *"shell"* plays an important role in the modulation of the structure and morphology of Fe-based species during the pyrolysis process. Indeed, as the precursor is heated to the pyrolysis temperature, it expels most of its oxygen atoms as volatile oxygen-based species such as water, $CO_2$ and CO [38] which, in principle, could react with the Fe to yield the $Fe_xO_y$ derivatives. In $FeFe_2$-$CN_l$ 900/C the relatively high N content in the *"coordination nests"* enhances the interactions between the CN matrix (the EC *"shell"*) and the Fe-based NPs, thus inhibiting the formation of $Fe_xO_y$ NPs. In $Fe_2Fe$-$CN_l$ 900/C the opposite situation is observed, demonstrating that a lower N content (see Table 1) reduces the density and strength of metal-N interactions in *"coordination nests"*, facilitating the aggregation and growth of $Fe_xO_y$ NPs such as $Fe_3O_4$ (see Table 2). Furthermore, the XRD patterns indicate that **A** process influences significantly the structure of the CN-based ECs, as follows: (1) for $FeFe_2$-$CN_l$ 900/$C_A$, a reduction of the overall abundance of Fe-based phases is observed; in this case, traces of cohenite ($Fe_3C$, COD#1010936 [56]), a close analogue of cementite, together with a small amount of austenite, (γ-Fe, COD#9012188 [56]), are detected; (2) for $Fe_2Fe$-$CN_l$ 900/$C_A$, no Fe-based phase is detected (see Figure 2(d) and Table 2). These pieces of evidence are in accordance with the chemical composition results (see Table 1), demonstrating that the interactions between the CN matrix (the *"shell"*) and Fe-based species are crucial in modulating the structure and the morphology of active sites in these ECs. In details, in $FeFe_2$-$CN_l$ 900/C, the strong [Fe-based NP]/[CN matrix] interactions inhibit the complete elimination of Fe-based phases upon **A**. The opposite behavior is revealed for $Fe_2Fe$-$CN_l$ 900/C where, owing to the weak [Fe-based NP]/[CN matrix] interactions, no Fe-based phases are left in the ECs after **A**.

### 3.4. Morphology

FIGURE 3

In SEM micrographs (see Figure 3) of proposed ECs a dark background is visible, corresponding to the carbon-based matrix, which is dotted by bright submicrometric (d < 200 nm) spots representative of metal-rich grains. A comparison of the morphology of pristine ECs (Figure 3(a) and Figure 3(b)) with that of activated ECs (Figure 3(c) and Figure 3(d)) highlights a significant rarefaction of the bright spots. In particular, no bright grains are revealed on $Fe_2Fe$-$CN_l$ 900/$C_A$. This evidence demonstrates that **A** is a very efficient process in eliminating the Fe-based NPs in this sample (see Table 1 and Figure 2(d)). Figure 4 displays the HR-TEM images of the CN-based ECs.

FIGURE 4

A careful analysis of the micrographs of Figure 4(a) and Figure 4(c) shows that the pristine CN-based ECs consist of aggregates of *"core-shell"* NPs. It is to be noticed that, interestingly: (i) the XC-72R carbon black *"core"* NPs are uniformly covered by a foamy carbon nitride *"shell"* embedding the dark metal-rich nanoparticles with sizes in the range 20 < d < 80 nm; and (ii) **A** process is very efficient to disaggregate the *"core-shell"* nanoparticles of the ECs and eliminate Fe-based NPs from both $FeFe_2$-$CN_l$ 900/$C_A$ and $Fe_2Fe$-$CN_l$ 900/$C_A$ (see Figure 4(b) and Figure 4(d)). The high-resolution TEM images of $FeFe_2$-$CN_l$ 900/C and $FeFe_2$-$CN_l$ 900/$C_A$ (see Figure 5) show that **A** does not affect the morphology of the CN "shell" significantly.

FIGURE 5

The HR-TEM micrographs of Figure 5 provide crucial information on the morphology of the CN *"shell"* matrix and on the chemical composition (see Table 1) of ECs, which corroborates and completes the structural studies carried out by powder XRD (see Section 3.3). In detail, the interplanar distances measured in the micrographs of Figure 5 coincide with powder XRD results and prove that the following metal NPs are present: (a) for $FeFe_2$-$CN_l$ 900/C, systems with an interplanar distance of *ca.* 2.4 Å (see Figure 5(a)), corresponding to the typical $d_{121}$ values of cementite ($Fe_3C$); (b) for $FeFe_2$-$CN_l$ 900/$C_A$, the revealed interplanar distance is *ca.* 2.1 Å (see Figure 5(b)), which is attributed to the $d_{111}$ of fcc γ-Fe (austenite). The presence of these NPs induces in both ECs ordering in the CN *"shell"*. Indeed, well-ordered *"onion-like"* carbon nitride layers are formed, which encapsulate the metal NPs in *"coordination nests"*. The CN *"shell"* presents the typical $d_{002} \approx 3.6$ Å interplanar distance usually revealed in graphite-like carbon nitride structures [38, 44]. It should be noted that upon **A**: (a) the thickness of the *"onion-like"* CN *"shell"* is reduced by 50% (from ≈ 6 to ≈ 3 nm for $FeFe_2$-$CN_l$ 900/C and $FeFe_2$-$CN_l$ 900/$C_A$, respectively); and (b) the overall porosity of the ECs rises, as revealed by HR-TEM micrographs (see Figure 4(a) and Figure 4(b)). **A** also acts to etch Fe species significantly (see Table 1); accordingly, the size of metal NPs is decreased (see Figure 4(a) and Figure 4(b)). However, in $FeFe_2$-$CN_l$ 900/$C_A$ the surface of metal NPs is still wrapped by the *"onion-like"* CN layers (see Figure 5(b)). This evidence points out that the latter layers are very flexible. Indeed, they are able to shrink, matching the morphology of the underlying metal NP template.

A similar behavior of morphology is revealed for pristine $Fe_2Fe$-$CN_l$ 900/C, where $Fe_3O_4$ (magnetite) NPs with an interplanar distance of $d_{131} \approx 2.5$ Å, are covered by a CN *"shell"*. The inspection of HR-TEM images of $Fe_2Fe$-$CN_l$ 900/C and $Fe_2Fe$-$CN_l$ 900/$C_A$ shows that: (a) the structure and composition of NPs are in accordance with powder XRD studies (see Table 2); (b) a clear *"core-shell"* morphology characterizes the NPs; and (c) the CN *"shell"* shows a foamy morphology and wraps the XC-72R *"core"* NPs (see Figure S2 of Supplementary Information).

TABLE 3

A comparison between the results reported in Table 3 indicates that **A** raises the BET area of the ECs. It should be pointed out that in both activated CN-based ECs, the area is relatively close to that of XC-72R support NPs. Taken all together, results allow to conclude that **A** is a very effective process: (a) for the disaggregation of the *"core-shell"* NPs; and (b) in the reduction of the thickness of their CN *"shell"*.

**3.5. X-ray photoelectron spectroscopy studies**

Further information on the correlation between the structure and the performance of the proposed ECs is obtained by investigating the surface chemical features of CN-based ECs and XC-72R by XPS.

TABLE 4

Table 4 indicates that the surface concentration of Fe: (a) decreases in the order $Fe_2Fe$-$CN_l$ 900/C > $FeFe_2$-$CN_l$ 900/C; and (b) is lower than the detection limit of the XPS technique for activated CN-based ECs. This, in accordance with the morphological and structural information previously discussed, shows that Fe-based NPs: (a) are encapsulated in compact *"onion-like"* CN *"shells"* in $FeFe_2$-$CN_l$ 900/C (see Figure 5(a)), thus compromising their exposure to the outer surface of the CN-based *"shells"*; (b) in $Fe_2Fe$-$CN_l$ 900/C, they are embedded in a very porous and foam-like carbon nitride *"shell"* (see Figure S2 in Supplementary Information), which facilitates their exposure on the external EC surface. **A**, as expected, is very effective in the elimination of Fe species (see Table 1, Figure 3 and Figure 4). Indeed, after **A** only Fe complexes strongly bound to the CN matrix are selectively left in the *"shell"*, thus yielding very stable active sites stabilized by the *"coordination nests"*. The lack of detection of these Fe complexes by XPS confirms these results. Furthermore, with respect to XC-72R, in CN-based ECs the concentration of O species is lower (see Table 4). This suggests that **A** promotes the selective elimination of O species, improving the graphitization of the CN matrix. In $Fe_2Fe$-$CN_l$ 900/$C_A$ O species are eliminated concurrently with the surface $Fe_xO_y$ inorganic species. The N density in the CN matrix increases on the amount of stoichiometric nitrogen of the reagents: $Fe_2Fe$-$CN_l$ 900/C < $FeFe_2$-$CN_l$ 900/C. Both $FeFe_2$-$CN_l$ 900/$C_A$ and $Fe_2Fe$-$CN_l$ 900/$C_A$ (see Section 3.3. and Section 3.4.): (a) exhibit a very rough and porous morphology; and (b) have lost most of their Fe-rich phases. In the latter case, only bulk Fe-complexes strongly bound in N *"coordination nests"* of CN matrix are present which results in the absence of Fe signal in limited to the top surface XPS data.

FIGURE 6

The curve fit of the high-resolution XPS N1s region (see Figure 6) shows six peaks, which are assigned to the following six different N-functionalities [52, 57, 58]: cyano or imine (398 eV); pyridinic N (398.6 eV); $N_x$-Fe species (399.6 eV); pyrrolic N (400.7 eV); graphitic N with possible contribution of quaternary N (401.8 eV); and oxidized nitrogen species (403 eV). A comparison of the spectral features of these peaks reveals that after **A**: (i) the intensities of oxidized N peak (see Figure 6) decrease, confirming that the graphitization of the CN matrix of the EC *"shells"* is improved; (ii) the density of N-functionalities of pyrrolic nitrogen increased also due to increased graphitization; and (iii) the pyridinic and N-ligand functionalities involved in the $N_x$-Fe *"coordination nests"*, which are expected to play a crucial role in the ORR process[58, 59], are only slightly affected. Taken all together, results allow us to summarize that: (a) the relative intensity of the peak ascribed to pyrrolic-like N functionalities increases in the order: $FeFe_2$-$CN_l$ 900/C < $FeFe_2$-$CN_l$ 900/$C_A$ < $Fe_2Fe$-$CN_l$ 900/C ≈ $Fe_2Fe$-$CN_l$ 900/$C_A$; and (b) the peak intensity ratio $N_x$-Fe/pyrrolic N in $FeFe_2$-$CN_l$ 900/C and $FeFe_2$-$CN_l$ 900/$C_A$ is larger than the values of $Fe_2Fe$-$CN_l$ 900/C and $Fe_2Fe$-$CN_l$ 900/$C_A$ ECs (see Table S1). This permits us to propose that probably: (a) pyrrolic-like N functionalities act to facilitate the first 2-electron reduction step of $O_2$ into $H_2O_2$ [52, 58, 59]; and (b) $N_x$-Fe species play a crucial role in the modulation of the overall ORR kinetics in *"Pt-free"* ECs [58, 60]. Now, if we consider that the density of pyrrolic N-like functionalities and $N_x$-Fe coordination species are crucial parameters for the modulation of the ORR activity and selectivity of ECs, it is expected that the best EC in the ORR process is $FeFe_2$-$CN_l$ 900/$C_A$. The latter EC has a higher relative amount of N coordinated to iron with respect to pyrrolic N, that facilitates more efficient reduction of hydrogen peroxide by $N_x$-Fe species. This sample also includes the $Fe_3C$ phase which, according to *"in-situ"* XAS (X-ray Absorption Spectroscopy) studies, should be inactive towards $O_2$ adsorption; on the other hand, subsurface Fe/$Fe_xC$ may also act to stabilize the peroxide intermediate on the active site in the alkaline medium. [61]

### 3.7. CV-TF-RRDE investigations

The *"ex situ"* performance of the CN-based ECs in the ORR process is investigated by CV-TF-RRDE in order to detect the kinetic parameters describing the reduction mechanisms and the selectivity towards the four-electron process. The ORR disk current densities ($J_{ORR}$) and ring currents ($I_r$) of both pristine and activated ECs on E *vs.* the reversible hydrogen electrode (RHE) are shown in Figure 7.

FIGURE 7

The Tafel plots derived from the ORR profiles reported in Figure 7 [44] are shown in Figure 8.

FIGURE 8

In the acid environment the ORR Tafel slopes decrease from *ca.* 170 $mV \cdot dec^{-1}$ to *ca.* 120 $mV \cdot dec^{-1}$ upon **A**. In the alkaline environment, the Tafel slopes: (a) show values of *ca.* 70 $mV \cdot dec^{-1}$ at low ORR overpotentials ($\eta$); (b) increase on $\eta$; and (c) decrease after **A**. The specific surface activity (SSA) values of the ECs in the alkaline environment is obtained by means of Eq. 1:

$$SSA = \frac{i_k}{m_{EC} \cdot A_{BET}} \qquad \text{[Eq. 1]}$$

where $i_k$ is the ORR kinetic current determined on Figure 8(b) at 0.85 V *vs.* RHE, $m_{EC}$ is the EC mass deposited on the RRDE tip and $A_{BET}$ is the BET area of the ECs obtained by nitrogen physisorption technique (see Table 3). SSA results are summarized in Table 5.

TABLE 5

The selectivity in the 4-electron pathway of the ORR is gauged by evaluating the fraction of hydrogen peroxide ($X_{H_2O_2}$) oxidated at the ring by means of Eq. 2 [49]:

$$X_{\mathrm{H_2O_2}} = \frac{\frac{2 I_{\mathrm{r}}}{N}}{I_{\mathrm{D}} + \frac{I_{\mathrm{r}}}{N}} \qquad \text{[Eq. 2]}$$

where $I_D$ and $I_r$ are the current values measured concurrently on the glassy-carbon disk and on the platinum ring. The collection efficiency of the ring is N = 0.38. $X_{H_2O_2}$ results on E are plotted in Figure 9.

FIGURE 9

The selectivity in the 4-electron pathway of the ECs: (a) is significantly lower than that of the Pt/C ref.; (b) increases after **A** in the acid solution; and (c) shows the opposite trend of the acid solution in the alkaline environment. A quantitative comparison of the ORR performance of the proposed ECs is carried out evaluating the following figures of merit: (a) $E(J_{5\%})$, the *"onset potential"*; and (b) $X^*_{H_2O_2}$, the *"ORR selectivity at 0.3 V vs. RHE"*. $E(J_{5\%})$ is the electrode potential corresponding to an ORR current density equal to 5% of the maximum ORR limiting current determined for the Pt/C ref.

at *ca.* 0.3 V *vs.* RHE in the same conditions. $X^*_{H_2O_2}$ is obtained by Eq. 2 and the data of Figure 7 at a potential of 0.3 V *vs.* RHE.

FIGURE 10

Results (see Figures 7-10 and Table 5) show that in the acid environment $E(J_{5\%})$ values of the ECs are *ca.* 220 mV higher than that of Pt/C ref. Taking all together, this demonstrates that, in the alkaline environment, the overall ORR performance of the CN-based ECs is significantly improved (see Figure 7) and approaches that of the Pt/C ref. This behavior can be rationalized considering that: (a) in the acid environment, the ORR occurs when $O_2$ is adsorbed on the *"inner-sphere"* of EC NPs [62] which, for the Pt/C ref., yields strong Pt-$O_{2,ad}^-$ and Pt-$O_{ad}$ coordination bonds [63]; while (b) in the alkaline solutions, in accordance with other studies [62], the ORR takes place involving two concurring $O_2$ adsorption phenomena on the metal active sites. In this latter case, the two processes consist of: (i) a first one, similar to that occurring in acid solutions (α); and (ii) a second one, ascribed to a weaker *"outer-sphere"* $O_2$ adsorption event (β). (β), which is less modulated by the type of the active site, occurs quite easily even in the absence of active sites based on platinum-group metals (PGMs) [62]. In the alkaline environment, it is expected that for the Pt/C ref. both (α) and (β) processes are operating simultaneously, while in CN-based ECs (α) is inhibited and at low η the ORR proceeds mostly through (β) process. Now, if we admit that (β) is almost independent of the type of specific adsorption site, it is reasonable to assume that the reactivity of the entire surface of the CN-based ECs is playing a crucial role in the ORR process. A different behavior is detected in acid solutions where the ORR process, which is modulated by the *"inner-shell"* $O_2$ adsorption phenomena, occurs on the surface of the active metal sites. On this basis, SSA (see Eq. 1 and Table 5) is a meaningful figure of merit to diagnose in the kinetic regime the performance of the CN-based ECs in alkaline solution. In detail, the SSA of the CN-based ECs is almost three orders of magnitude lower than that of the Pt/C ref. (see Table 5). Accordingly, with respect to (β) process, which occurs for both the Pt/C ref. and the CN-based ECs, (α) process, which takes place at low ORR overpotentials on the Pt sites of the Pt/C ref., is responsible for the main contribution to the significantly improved ORR kinetics in this case. Finally, at pH > 12, (β) yields a 2-electron exchange process, which produces an $HO_2^-$ intermediate [64]. $HO_2^-$ is reduced to water on the electrocatalyst surface, either: (i) on the same active site, or (ii) upon diffusion to another active metal site yielding a consecutive *"series"* of $2e^-$ x $2e^-$ processes with an overall $4e^-$ pathway [62].

Pristine $FeFe_2$-$CN_l$ 900/C shows the highest ORR η, both in the acid and in the alkaline environment (see Figure 7, and Figure 8). This result is consistent with a significant aggregation of the EC nanoparticles, yielding the lowest overall BET area (see Table 3) where the ORR can take place. In the acid environment, the Tafel slope of pristine $FeFe_2$-$CN_l$ 900/C is very high, *ca.* 170 $mV \cdot dec^{-1}$ (see Figure 8). This value is associated with a very low intrinsic ORR kinetics [52]. In the alkaline environment, the Tafel slope of pristine $FeFe_2$-$CN_l$ 900/C: (i) at low ORR η is equal to *ca.* 80 $mV \cdot dec^{-1}$; and (ii) increases up to more than 150 $mV \cdot dec^{-1}$ as the ORR η is raised. This can be attributed to a progressive modulation *vs.* η of the $O_2$ adsorption isotherm [50]. Indeed, at low η, $O_2$ adsorption likely takes place owing to a mixed Temkin-Langmuir process with a Tafel slope ranging between 60 (*"pure"* Temkin isotherm) and 120 $mV \cdot dec^{-1}$ (*"pure"* Langmuir isotherm) [65]. This phenomenon is explained if we consider that the first electron transfer process, which takes place immediately after the *"outer-sphere"* $O_2$ adsorption event, becomes progressively easier and faster as the electron density at the electrode rises. In this way: (i) the overall ORR kinetics is improved; and (ii) the adsorption sites are regenerated with a higher rate. In the alkaline solutions, $X^*_{H_2O_2}$ of pristine $FeFe_2$-$CN_l$ 900/C is the lowest (16.4%), since the aggregation of the CN-based EC NPs inhibits the elimination of the $HO_2^-$ intermediate from the electrode layer thus reducing its detection on the ring electrode [66, 67]. On the contrary, in the acid environment, $X^*_{H_2O_2}$ of pristine $FeFe_2$-$CN_l$ 900/C is very high (*ca.* 35%), suggesting that its surface is very rich in oxygen-based species (see Table 4). Therefore, the dissociative adsorption of $O_2$ in *"inner sphere"* processes, which requires the presence of at least two adjacent free active sites [49, 68], is hindered. Upon **A**, the ORR performance of $FeFe_2$-$CN_l$ 900/C is significantly improved both in the acid and in the alkaline solutions (see Figure 7). This behavior is attributed to the ability of **A** to disaggregate the NPs of the ECs, raising both the BET area (see Table 3) and the number of ORR active sites. In the alkaline environment, on the basis of the SSA increase upon **A** (see Table 5) we can also assume that the turnover frequency of the active sites has been increased. Indeed, in this case, an efficient stabilization of the $HO_2^-$ ORR intermediate is expected due to the presence of active metal sites with a high concentration of cationic Fe coordination species, which are formed when carbon and nitrogen heteroatom-based ligands are present on the CN matrix (the *"coordination nests"*) [62]. In the alkaline environment, **A** improves the kinetics of the ORR processes and reduces $\eta$: $E(J_{5\%})$ increases from 0.855 to 0.908 V *vs.* RHE. Consequently, the contribution of $O_2$ adsorption process following the Temkin isotherm is increased, and the Tafel slopes are concurrently reduced (see Figure 8) [50]. After **A** the ORR overpotentials of ECs in the acid solutions are very significant ($E(J_{5\%})$ < 750 mV *vs.* RHE). This result suggests that no $O_2$-based intermediates are adsorbed on the surface of the active sites, as expected on the basis of a Langmuir-like isotherm with a Tafel slope equal to *ca.* 120 $mV \cdot dec^{-1}$. The elimination of O-based species from the active sites facilitates the dissociative adsorption of $O_2$, reducing $X^*_{H_2O_2}$ in the acid environment from 35.2% for $FeFe_2$-$CN_l$ 900/C to 25.9% for $FeFe_2$-$CN_l$ 900/$C_A$ (see Figure 9 and Figure 10). In the alkaline solution, after **A** the disaggregation of the EC NPs facilitates the elimination of the $HO_2^-$ intermediates from the active site surface. The intermediates then form $H_2O_2$, which is easily detected at the ring electrode. The latter reveals that $X^*_{H_2O_2}$ increases from 16.4% for $FeFe_2$-$CN_l$ 900/C to 34.0% for $FeFe_2$-$CN_l$ 900/$C_A$ (see Figure 9 and Figure 10).

A comparison of the ORR profiles (see Figure 7) shows that the ORR performance improves in the order $FeFe_2$-$CN_l$ 900/C < $Fe_2Fe$-$CN_l$ 900/C. It is observed that for $Fe_2Fe$-$CN_l$ 900/C: (i) in the alkaline environment, $E(J_{5\%})$ > *ca.* 30 mV (0.887 and 0.855 mV *vs.* RHE for $Fe_2Fe$-$CN_l$ 900/C and $FeFe_2$-$CN_l$ 900/C, respectively); and (ii) in the acid environment, $J_{ORR}$ > *ca.* 50% at V < 0.6 V *vs.* RHE (see Figure 7(a)). Similar Tafel slopes suggest that the ORR mechanism in pristine $FeFe_2$-$CN_l$ 900/C and $Fe_2Fe$-$CN_l$ 900/C is quite similar to one another for both the acid and the alkaline condition (see Figure 8). The improved ORR performance of pristine $Fe_2Fe$-$CN_l$ 900/C over $FeFe_2$-$CN_l$ 900/C can be attributed to the much larger BET area of $Fe_2Fe$-$CN_l$ 900/C (see Table 3), which comprises a much higher number of active sites. This interpretation is further supported by the analysis of $X^*_{H_2O_2}$ values which, in alkaline solutions, are higher for $Fe_2Fe$-$CN_l$ 900/C than for $FeFe_2$-$CN_l$ 900/C (20.8% *vs.* 16.4%, respectively, see Figure 10). Indeed, as the BET area of ECs is increased, the $OH_2^-$ intermediates are better eliminated from the electrode layer and easily detected at the RRDE ring. On the other hand, with respect to $FeFe_2$-$CN_l$ 900/C, the SSA of $Fe_2Fe$-$CN_l$ 900/C in the alkaline environment is slightly lower (see Table 5). This result is rationalized considering that, with respect to $FeFe_2$-$CN_l$ 900/C, the surface concentration of cationic Fe-species bound by the carbon and nitrogen-based functionalities of the *"coordination nests"* is lower in $Fe_2Fe$-$CN_l$ 900/C (see Section 3.5). Accordingly, the long-range electrostatic stabilization of the $HO_2^-$ ORR intermediates is inferior, thus hindering $O_2$ adsorption by means of *"outer-shell"* processes and consequently slowing the overall ORR kinetics.

**A** significantly improves the ORR performance of $Fe_2Fe$-$CN_l$ 900/$C_A$. Indeed, $J_{ORR}$ of $Fe_2Fe$-$CN_l$ 900/$C_A$ in the mixed kinetic-diffusion region is larger by *ca.* 80% and 30% in the acid and alkaline environments, respectively (see Figure 7). This is likely associated to an increase of the BET area (see Table 3), which originates ECs with a significantly increased number of ORR active metal sites. This interpretation is further supported by the growth of $X^*_{H_2O_2}$ values in the alkaline environment from 20.8% to 34.3% for $Fe_2Fe$-$CN_l$ 900/C and $Fe_2Fe$-$CN_l$ 900/$C_A$, respectively (see Figure 10). Thus, a higher BET area facilitates the elimination of the $HO_2^-$ ORR intermediates, which are then detected at the RRDE ring. In the acid environment, after **A**, $X^*_{H_2O_2}$ of $Fe_2Fe$-$CN_l$ 900/C decreases from 35.0% to 19.3%, concurrently with the reduction of the surface density of O-based functionalities (see Table 4). These latter, as expected, hinder the dissociative adsorption of $O_2$. Finally, Table 5 shows that the SSA of $Fe_2Fe$-$CN_l$ 900/C does not change significantly upon **A**. This demonstrates that, after **A**, no effects are detected on the functional features responsible for the ORR process in $Fe_2Fe$-$CN_l$ 900/C. These features include the surface density of active cationic Fe complexes bound in the *"coordination nests"* of the CN matrix.

## 4. Conclusions

A very reliable protocol is proposed to obtain two groups of *"core-shell"* ORR CN-based ECs without PGMs. The *"cores"* consist of XC-72R carbon black NPs, which are covered by a thin carbon nitride *"shell"* that embeds Fe species bound into the matrix through nitrogen and carbon-based *"coordination nests"*. In pristine ECs, the N wt% rises in the order $Fe_2Fe$-$CN_l$ 900/C < $FeFe_2$-$CN_l$ 900/C, concurrently with the thermal stability under inert $N_2$ atmosphere. The reverse trend is observed for the thermal stability in an oxidizing atmosphere, as witnessed by the main degradation event $T_D$ which decreases from 550°C to 400°C. $T_D$ is diagnostic of the capability of Fe species in active sites to promote the adsorption of $O_2$. HR-TEM shows that the CN *"shells"* embed $Fe_3C$ and $Fe_3O_4$ NPs. In $FeFe_2$-$CN_l$ 900/C, the $Fe_3C$ nanoparticles are encapsulated in well-ordered *"onion-like"* carbon nitride based *"coordination nests"*. It is demonstrated that **A** strongly affects the physicochemical properties of the ECs, yielding: (a) a significant increase in the degree of graphitization of the CN *"shell"*; (b) a substantial reduction in the wt% of the Fe-based phases; and (c) the disaggregation of the NPs, with a subsequent increase of the porosity and of the BET area of the ECs. The electrochemical results prove that **A** does not change significantly the electronic structure of Fe complexes in *"coordination nests"*, which modulates their performance in the ORR process. On the other hand, the Fe-based species not stabilized in the CN *"coordination nests"* are easily removed during **A** and are not expected to affect the ORR performance of the ECs significantly. CV-TF-RRDE shows that the best ORR performance of here proposed ECs is found in alkaline conditions where, with respect to the Pt/C ref., the ORR onset potential $E(J_{5\%})$ of $FeFe_2$-$CN_l$ 900/$C_A$ material is *ca.* 38 mV lower (0.946 and 0.908 V *vs.* RHE, respectively).

Taking all together, in accordance with other results [60], the ORR performance and selectivity are correlated with the physicochemical properties of the CN matrix (the *"shell"*) of here-proposed ECs. Furthermore, the increase of $E(J_{5\%})$ in the order $FeFe_2$-$CN_l$ 900/C < $Fe_2Fe$-$CN_l$ 900/C and after **A** treatment is consistent with the following phenomena: (i) the BET area of the ECs is increased (see Table 3); and (ii) the surface concentration of Fe-based complexes bound by nitrogen- and carbon-based *"coordination nests"* is raised. Indeed, after **A** the absolute number of active sites in ECs is increased, thus raising $X^*_{H_2O_2}$ in the alkaline environment from *ca.* 18% to *ca.* 35% respectively for pristine and activated ECs, respectively. These results suggest that the disaggregation of the EC NPs facilitates the elimination of $HO_2^-$ species from the EC surface owing to a concurrent multi-site mechanism as confirmed by the concentration of $H_2O_2$ detected at the RRDE ring. Finally, it should be pointed out that the presence of a high density of $FeC_xN_y$ complexes on the surface of the ECs improves the intrinsic ORR turnover frequency, which is also triggered by the coordination of the $HO_2^-$ ORR intermediates into stable Fe-based complexes strongly bound in *"coordination nests"* [62]. This phenomenon is particularly evident for $FeFe_2$-$CN_l$ 900/C and $FeFe_2$-$CN_l$ 900/$C_A$ ECs, where it is revealed a higher wt% of nitrogen ligands, which are stabilizing the Fe-complexes fixed in *"coordination nests"* of the CN matrix.

In conclusion, this report describes an efficient synthesis protocol to obtain *"Pt-free"* CN-based ECs with (a) a very good ORR performance in the alkaline environment; and (b) a highly promising perspective for application in energy conversion devices such as AEMFCs and metal-air batteries.

**Acknowledgements**

This work was funded by the Strategic Project of the University of Padova *"From Materials for Membrane-Electrode Assemblies to Energy Conversion and Storage Devices - MAESTRA"*. The research leading to these results has received funding from the European Union's Horizon 2020 research and innovation programme under grant agreement n°696656.

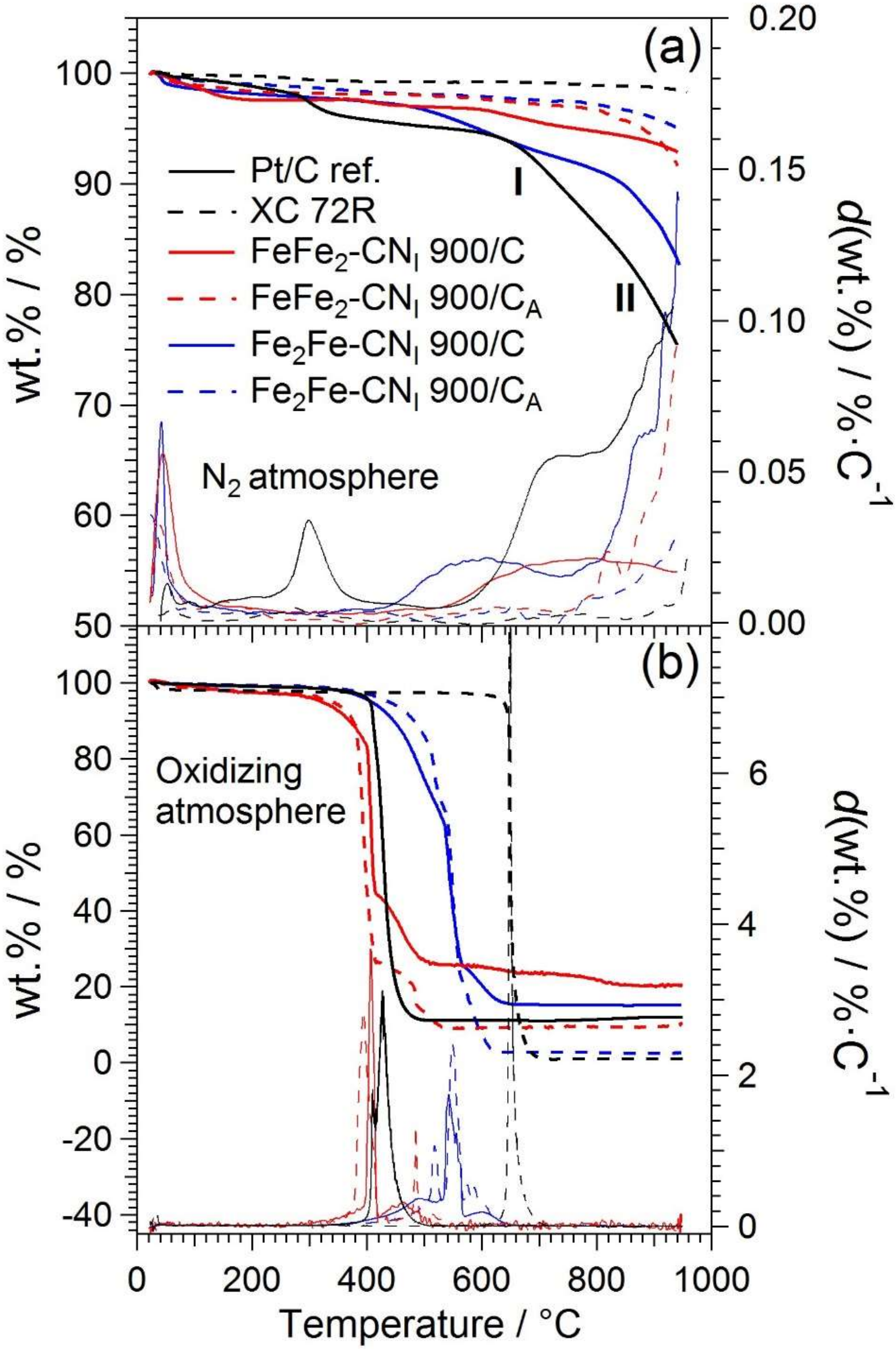

FIGURE 1

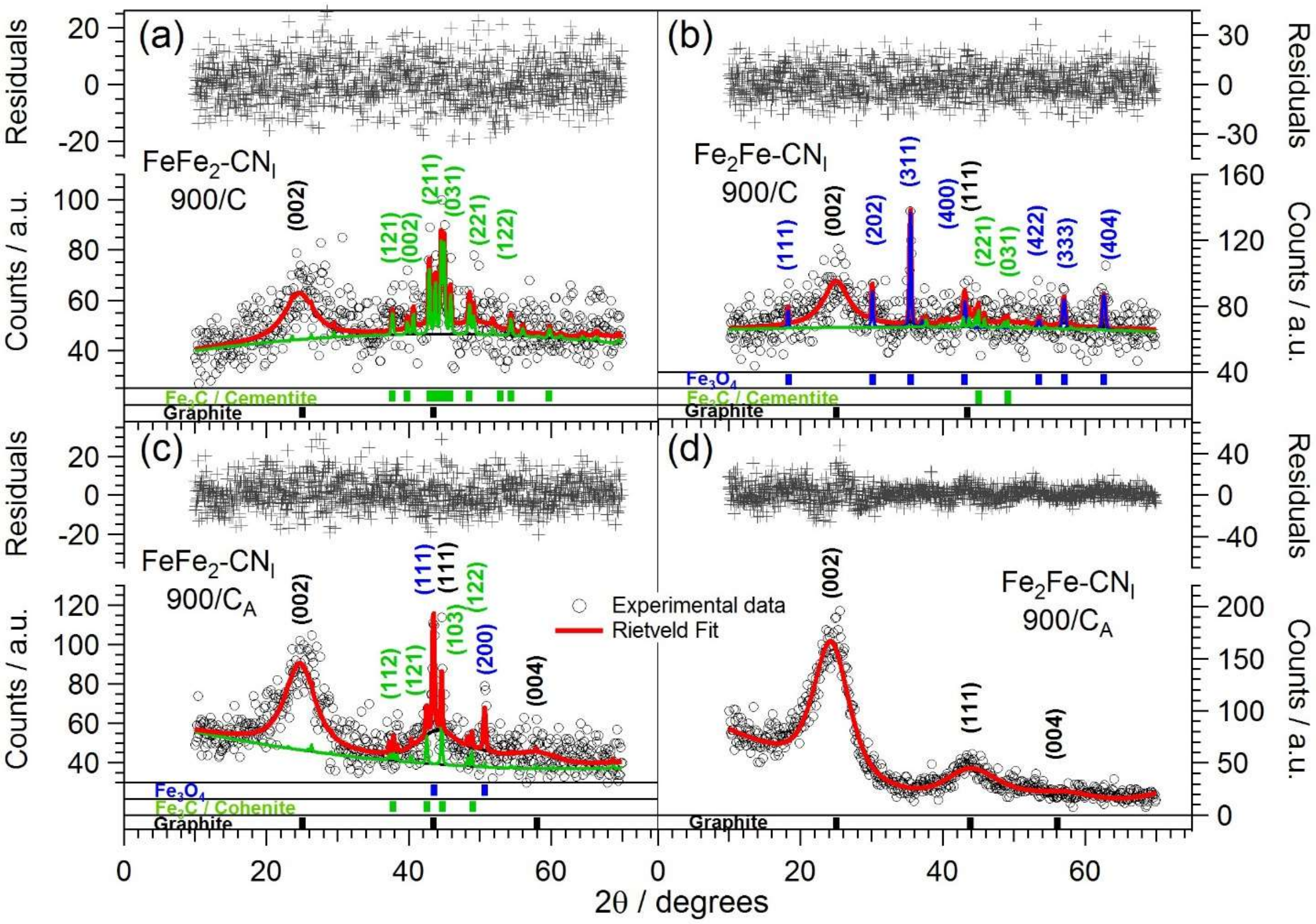

FIGURE 2

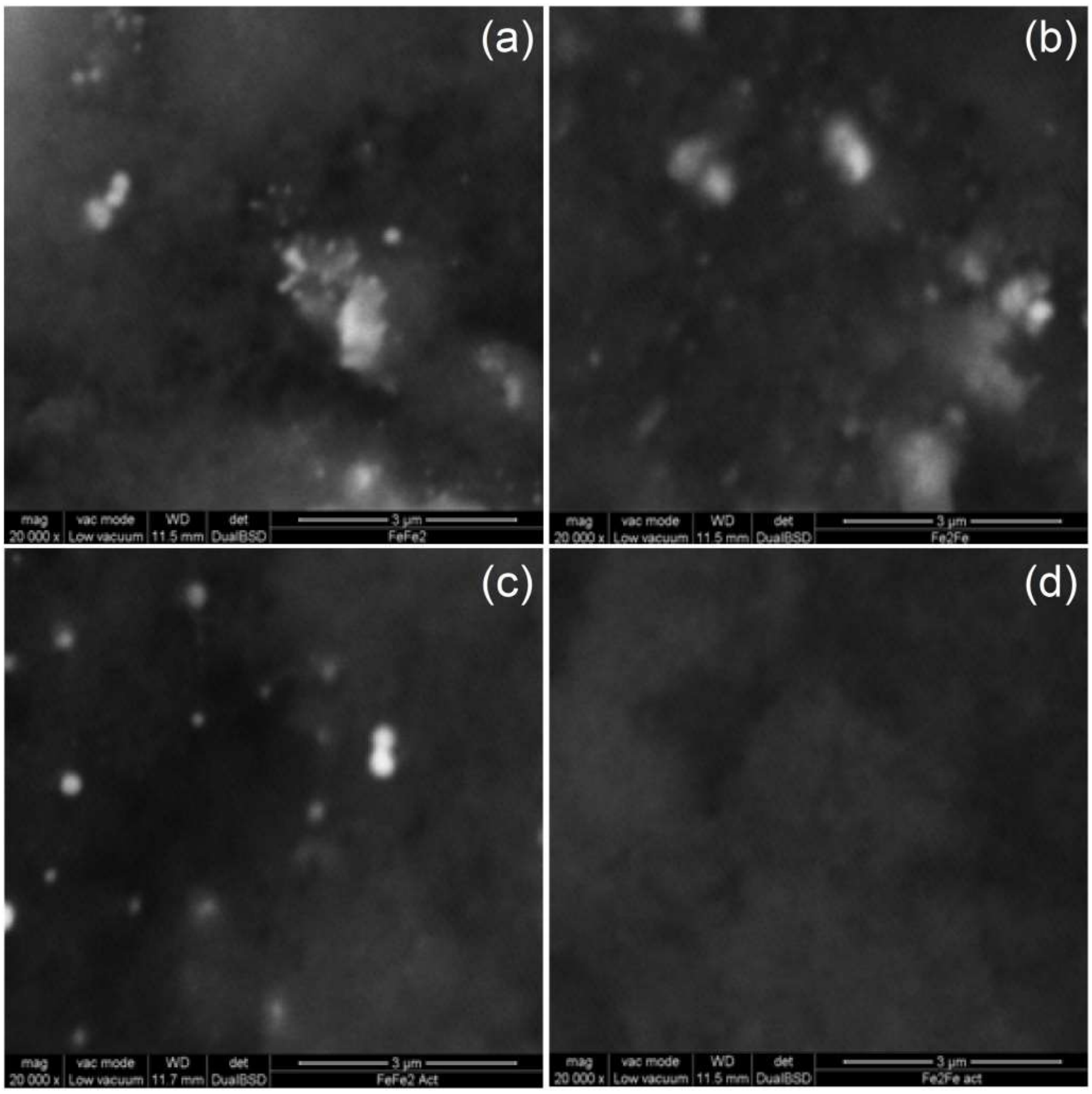

FIGURE 3

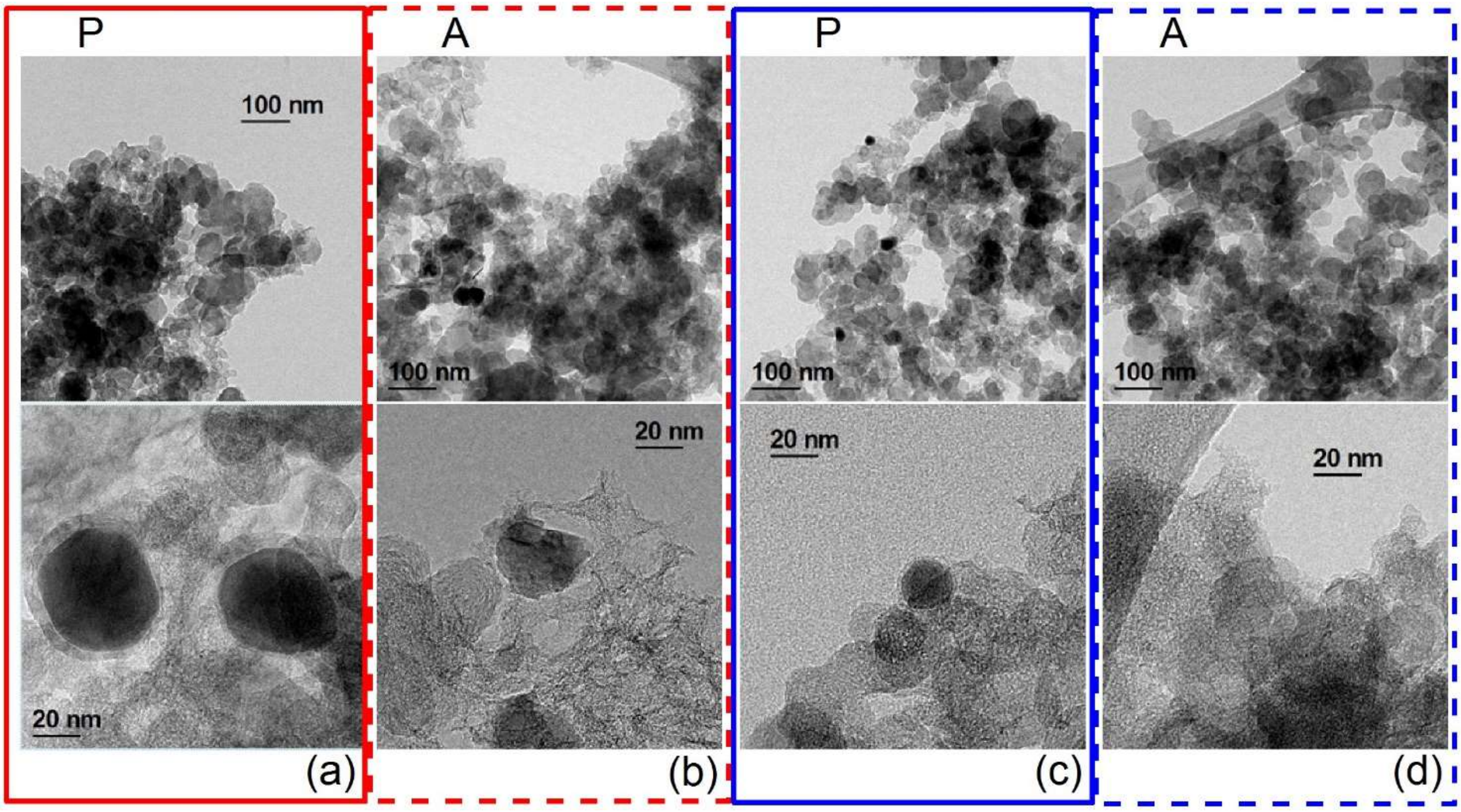

FIGURE 4

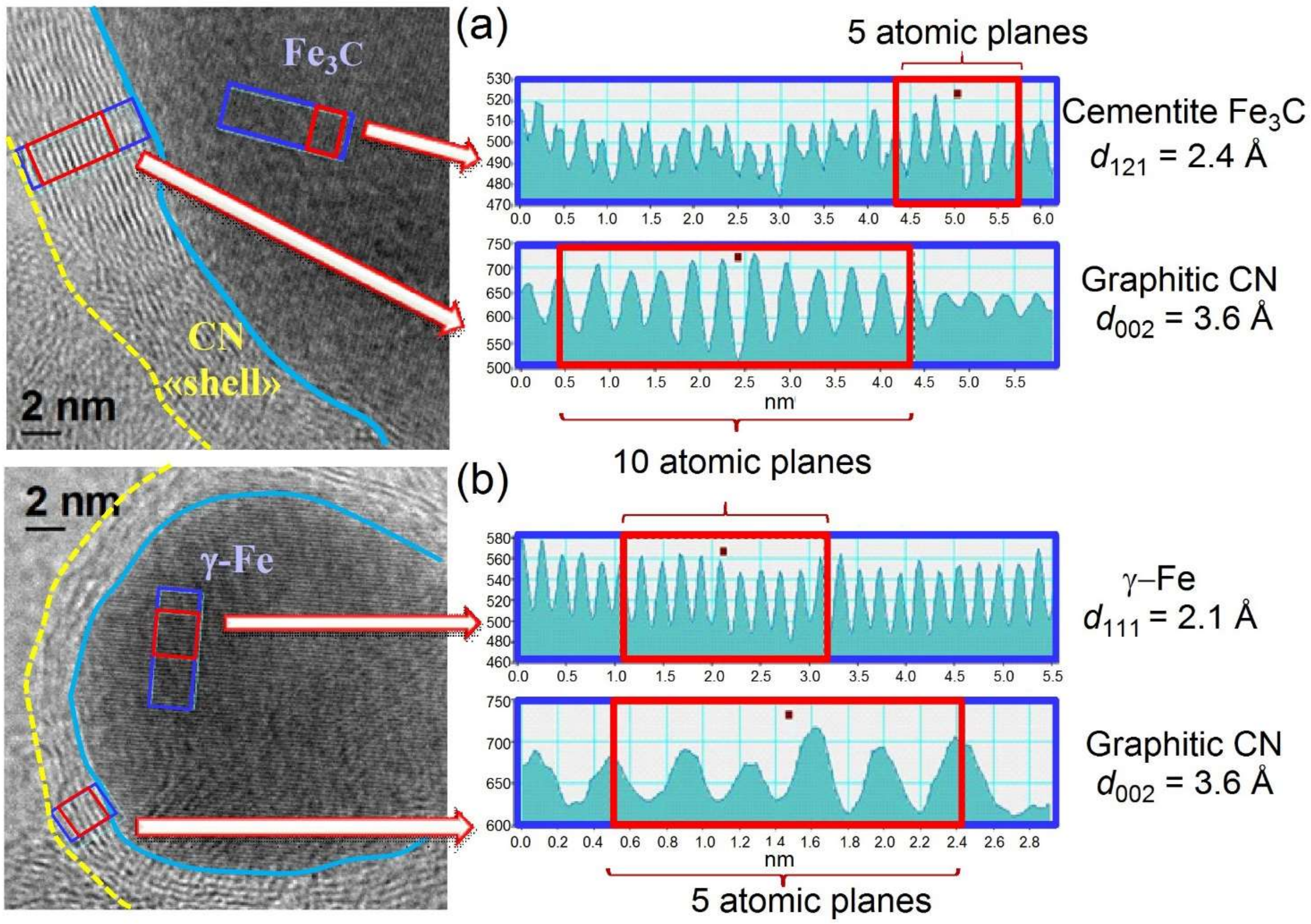

FIGURE 5

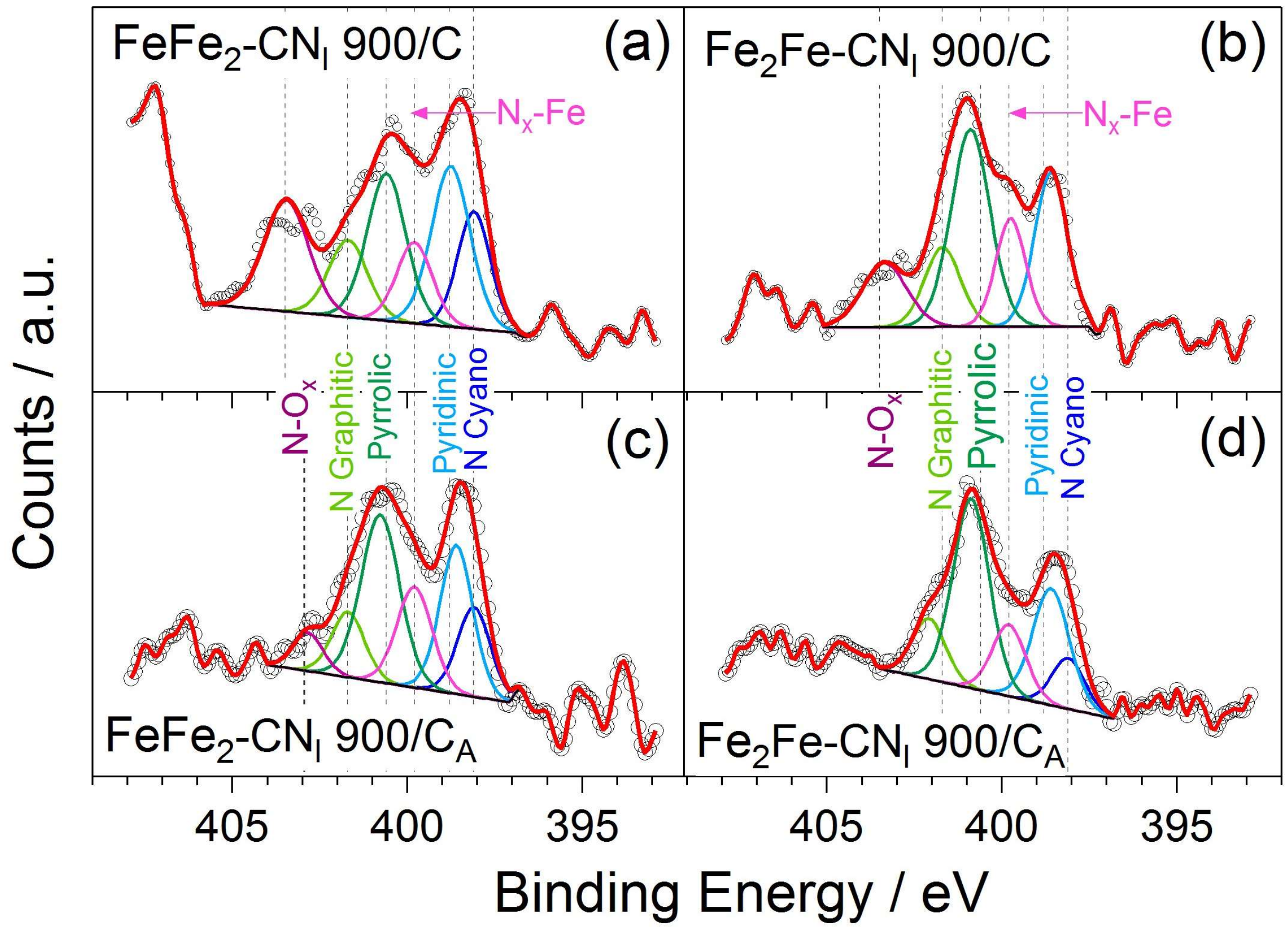

FIGURE 6

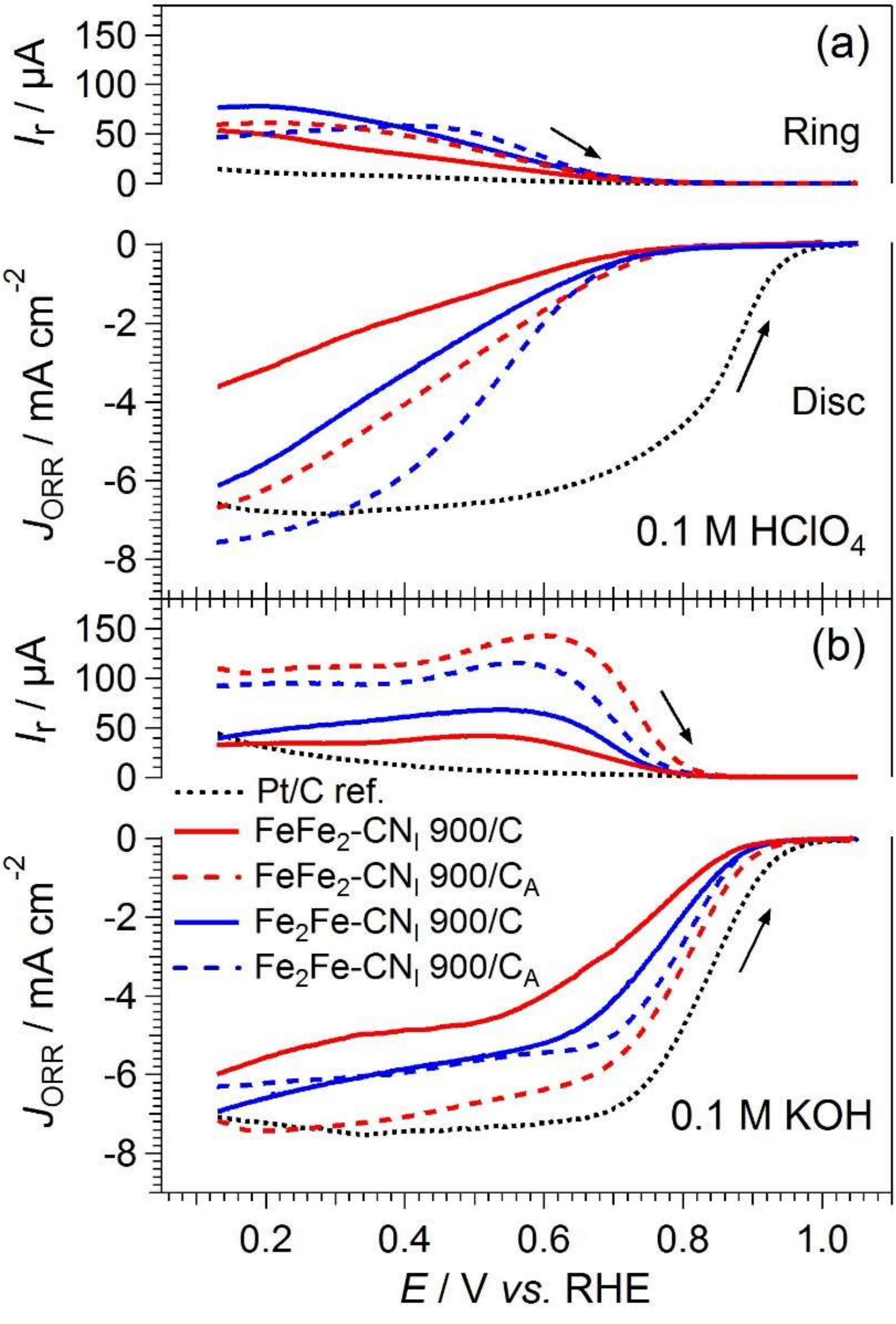

FIGURE 7

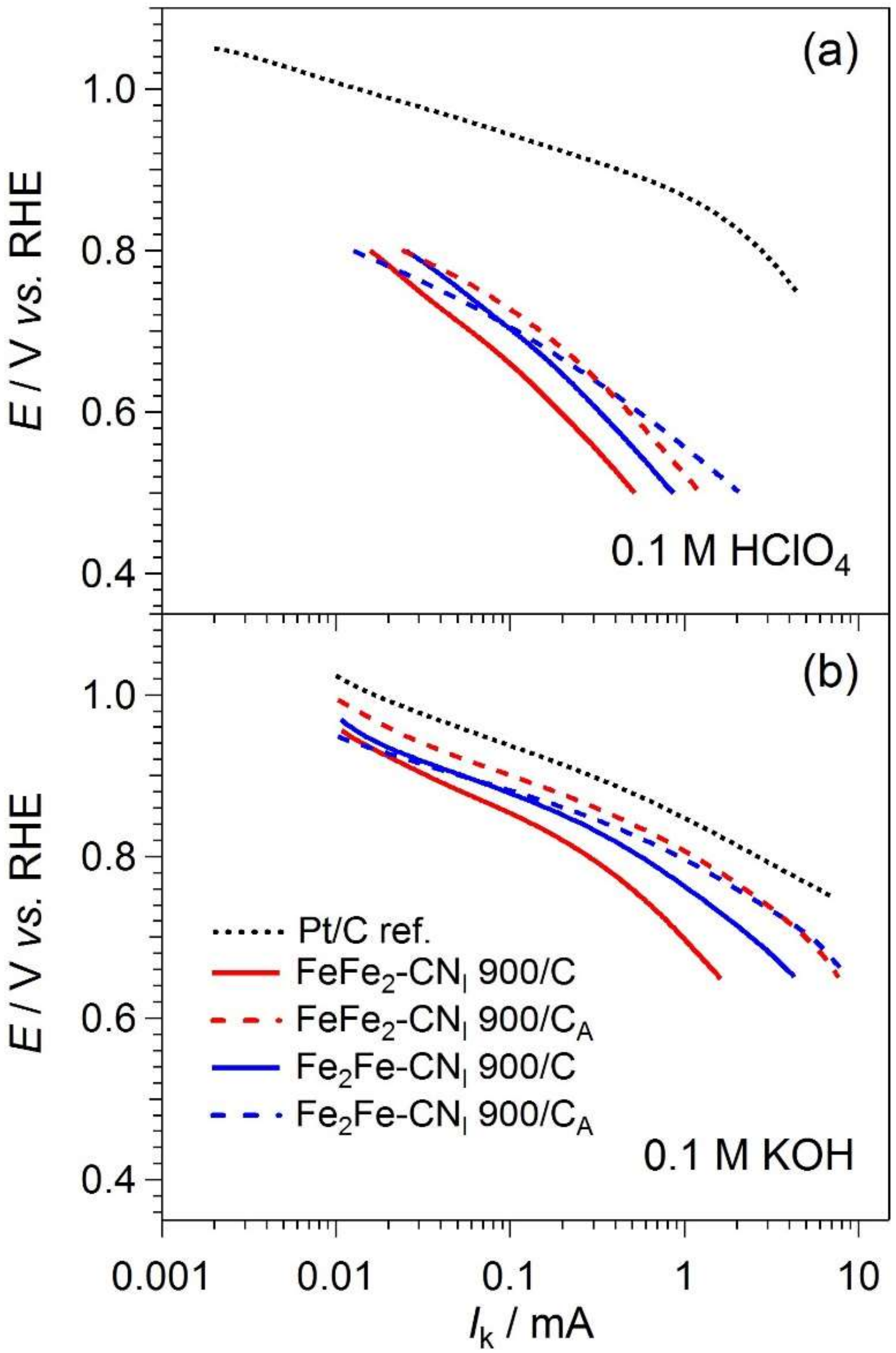

FIGURE 8

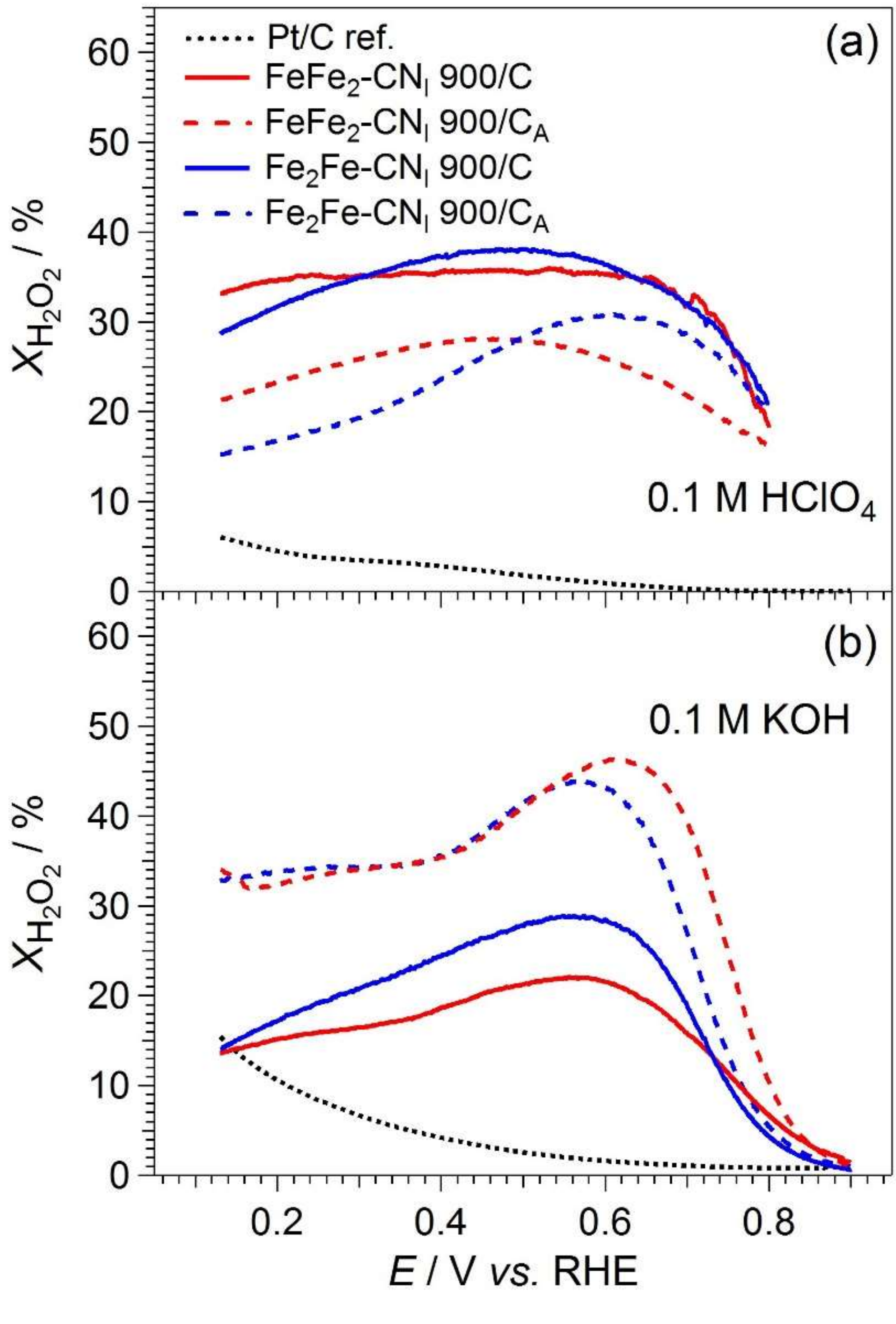

FIGURE 9

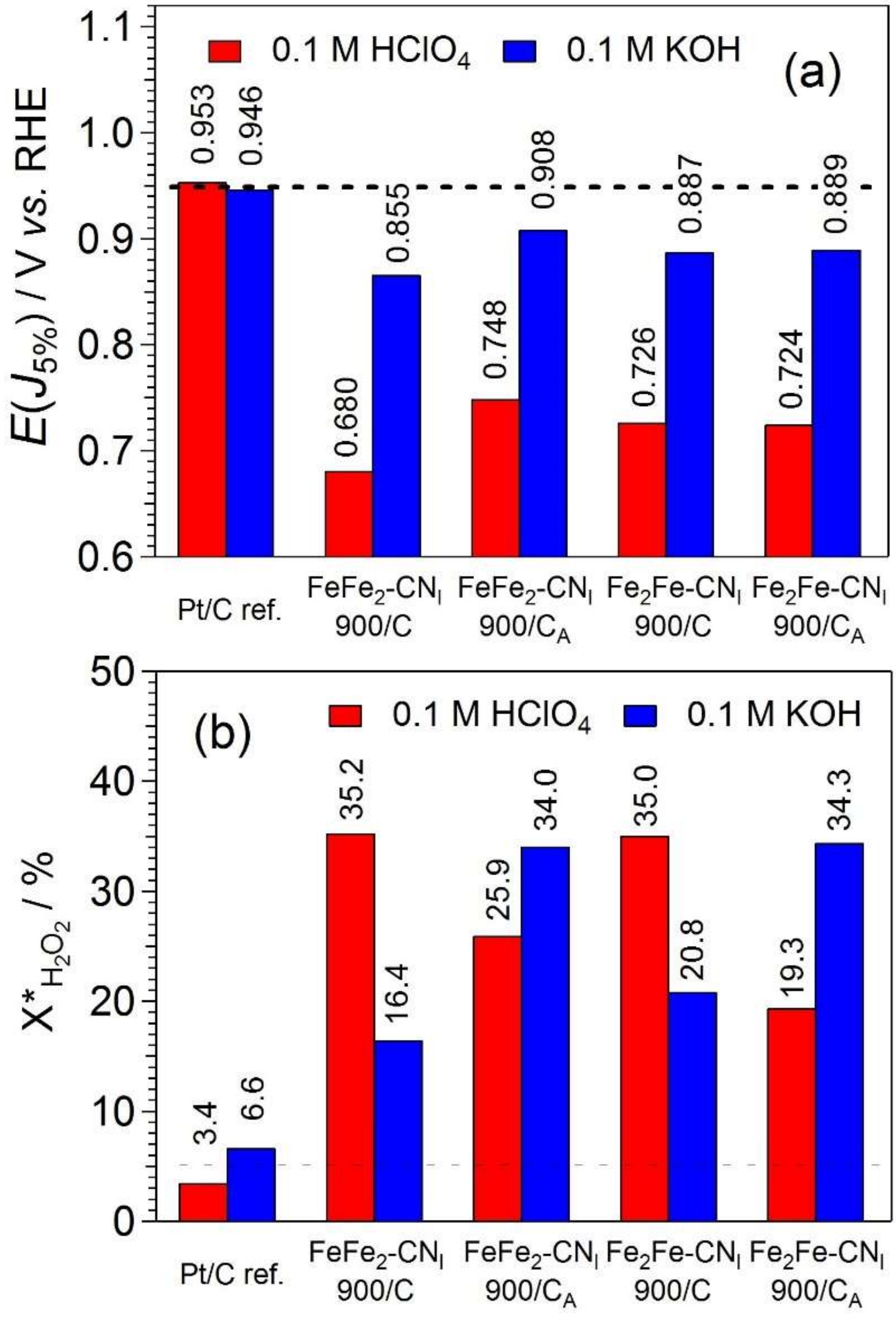

FIGURE 10

| Electrocatalyst | Weight / % | | | | | | Formula |
|---|---|---|---|---|---|---|---|
| | K(a) | Fe(a) | C(b) | H(b) | N(b) | S(b) | |
| ***FeFe₂-CN_l 900/C*** | 0.95 | 8.8 | 73.5 | 0.47 | 0.74 | 0.36 | $K_{0.15}[FeC_{38.8}H_{2.96}N_{0.34}S_{0.07}]$ |
| ***FeFe₂-CN_l 900/C_A*** | 0.77 | 3.3 | 88.9 | 0.28 | 0.33 | 0.34 | $K_{0.33}[FeC_{125}H_{4.70}N_{0.40}S_{0.18}]$ |
| ***Fe₂Fe-CN_l 900/C*** | 0.62 | 9.6 | 80.7 | 0.33 | 0.38 | 0.43 | $K_{0.09}[FeC_{39.1}H_{1.90}N_{0.16}S_{0.08}]$ |
| ***Fe₂Fe-CN_l 900/C_A*** | 0.44 | 0.17 | 92.1 | 0.2 | 0.43 | 0.58 | $K_{3.70}[FeC_{2519}H_{65.2}N_{10.1}S_{5.9}]$ |

(a) Determined by ICP-AES.
(b) Determined by microanalysis.

TABLE 1

| Sample | Phase | Space group | Phase abundance / wt% | Cell Parameters / Å | Particle size / nm | $R_{wp}$ (a)/ % |
|---|---|---|---|---|---|---|
| ***$FeFe_2$-$CN_l$ 900/C*** | C | Hexagonal *P*6₃/*mc* | 90 | a = 2.48, c = 7.22 | 2.2 | 15.1 |
| | Cementite $Fe_3C$ | Orthorombic *Pnma* | 10 | a = 5.09, b = 6.75, c = 4.53 | > 80-90 | |
| ***$FeFe_2$-$CN_l$ 900/$C_A$*** | C | Hexagonal *P*6₃/*mc* | 95 | a = 2.46, c = 7.18 | 2.3 | 14.0 |
| | Austenite (γ-Fe) | Cubic *Fm*-3*m* | 3 | a = 3.60 | > 80-90 | |
| | Cohenite $Fe_3C$ | Orthorombic *Pbnm* | 2 | a = 4.68, b = 5.08, c = 6.75 | > 80-90 | |
| ***$Fe_2Fe$-$CN_l$ 900/C*** | C | Hexagonal *P*6₃/*mc* | 85 | a = 2.52, c = 7.10 | 2-3 | 12.2 |
| | Magnetite $Fe_3O_4$ | Cubic *Fd*-3*m* | 12 | a = 8.38 | > 80-90 | |
| | Cementite $Fe_3C$ | Orthorombic *Pnma* | 3 | a = 5.09, b = 6.74, c = 4.53 | > 80-90 | |
| ***$Fe_2Fe$-$CN_l$ 900/$C_A$*** | C | Hexagonal *P*6₃/*mc* | 100 | a = 2.48, c = 7.29 | 2.5 | 15.9 |
| ***XC-72R*** | C | Hexagonal *P*6₃/*mc* | 100 | a = 2.48, c = 7.32 | 3.4 | 16.5 |

(a) $R_{wp}$ expresses the goodness of the fit; it is obtained by $R^2_{wp} = \Sigma_i w_i (y_{C,i}-y_{O,i})^2/\Sigma_i w_i (y_{O,i})^2$ where $w_i$ are the weights ($1/\sigma^2[y_{O,i}]$), $y_{C,i}$ and $y_{O,i}$ are values of computes and observed intensities [55].

TABLE 2

| Electrocatalyst | BET area[a] / $m^2 \cdot g^{-1}$ |
|---|---|
| ***$FeFe_2$-$CN_l$ 900/C*** | 82 |
| ***$FeFe_2$-$CN_l$ 900/$C_A$*** | 189 |
| ***$Fe_2Fe$-$CN_l$ 900/C*** | 197 |
| ***$Fe_2Fe$-$CN_l$ 900/$C_A$*** | 245 |
| ***XC-72R*** | 194 |

[a] Determined by the Brunauer-Emmett-Teller (BET) method.

TABLE 3

| Electrocatalyst | at.% | | | |
|---|---|---|---|---|
| | C | O | N | Fe |
| ***$FeFe_2$-$CN_l$ 900/C*** | 91.1 | 8.1 | 0.79 | 0.08 |
| ***$FeFe_2$-$CN_l$ 900/$C_A$*** | 92.7 | 6.9 | 0.37 | -[a] |
| ***$Fe_2Fe$-$CN_l$ 900/C*** | 91.8 | 7.2 | 0.53 | 0.44 |
| ***$Fe_2Fe$-$CN_l$ 900/$C_A$*** | 96.5 | 3.0 | 0.55 | -[a] |
| ***XC-72R*** | 89.6 | 10.4 | -[a] | -[a] |

[a] This value is lower than the detection limit of the XPS instrumentation.

TABLE 4

| Electrocatalyst | SSA / $\mu A \cdot cm^{-2}$ |
|---|---|
| ***$FeFe_2$-$CN_l$ 900/C*** | 0.87 |
| ***$FeFe_2$-$CN_l$ 900/$C_A$*** | 1.41 |
| ***$Fe_2Fe$-$CN_l$ 900/C*** | 0.70 |
| ***$Fe_2Fe$-$CN_l$ 900/$C_A$*** | 0.74 |
| ***Pt/C ref.*** | 473[a] |

[a] This value refers only to the Pt nanocrystals supported on the Pt/C ref; the loading of Pt on the RRDE tip is 15 $\mu g \cdot cm^{-2}$, the Pt specific surface area, as determined by CO stripping, is equal to 68 $m^2 \cdot g_{Pt}^{-1}$.

TABLE 5

FIGURE CAPTIONS

Figure 1. HR-TGA profiles and associated derivatives of the CN-based ECs under: $N_2$ inert atmosphere (a); and air oxidizing atmosphere (b).

Figure 2. Experimental powder XRD patterns (open circles) and calculated Rietveld profiles (red curves) of the ECs: $FeFe_2$-$CN_l$ 900/C (a); $Fe_2Fe$-$CN_l$ 900/C (b); $FeFe_2$-$CN_l$ 900/$C_A$ (c); and $Fe_2Fe$-$CN_l$ 900/$C_A$ (d).

Figure 3. SEM images of pristine ($FeFe_2$-$CN_l$ 900/C (a) and $Fe_2Fe$-$CN_l$ 900/C (b)) and activated CN-based ECs ($FeFe_2$-$CN_l$ 900/$C_A$ (c) and $Fe_2Fe$-$CN_l$ 900/$C_A$ (d)). Images are collected with backscattered electrons.

Figure 4. TEM images of the *"pristine"* (P) ($FeFe_2$-$CN_l$ 900/C (a) and $Fe_2Fe$-$CN_l$ 900/C (c)) and of their respective *"activated"* (A) CN-based ECs ($FeFe_2$-$CN_l$ 900/$C_A$ (b) and $Fe_2Fe$-$CN_l$ 900/$C_A$ (d)).

Figure 5. HR-TEM micrographs of: *"pristine"* $FeFe_2$-$CN_l$ 900/C (a); and *"activated"* $FeFe_2$-$CN_l$ 900/$C_A$ electrocatalysts (b).

Figure 6. High resolution N 1s XPS spectra of the CN matrix of ECs: $FeFe_2$-$CN_l$ 900/C (a), $Fe_2Fe$-$CN_l$ 900/C (b), $FeFe_2$-$CN_l$ 900/$C_A$ (c) and $Fe_2Fe$-$CN_l$ 900/$C_A$ (d).

Figure 7. ORR profiles of the CN-based ECs and of the Pt/C ref. in an $O_2$ atmosphere. The cell is filled with: 0.1 M $HClO_4$ (a); and 0.1 M KOH (b). $T$ = 298 K, sweep rate = 20 mV $s^{-1}$, electrode rotation rate 1600 rpm, $P_{O_2}$ = 1 atm.

Figure 8. ORR Tafel plots of data shown in Figure 7. Cell filled with: 0.1 M $HClO_4$ (a) and 0.1 M KOH (b). The experimental conditions are reported in the caption of Figure 7.

Figure 9. ORR selectivity profiles of the CN-based ECs and of the Pt/C ref. in a pure $O_2$ atmosphere. The experimental conditions are shown in the caption of Figure 7. Cell filled with 0.1 M $HClO_4$ (a) and 0.1 M KOH (b).

Figure 10. Figures of merit for the performance of the CN-based ECs. Onset potential, $E(J_{5\%})$ (a); ORR selectivity ($X^*_{H_2O_2}$) at 0.3 V *vs.* RHE (b).

TABLE CAPTIONS

Table 1. The chemical composition of the CN-based ECs.

Table 2. Results of Rietveld analysis of CN-based ECs and of XC-72R powder XRD patterns.

Table 3. BET area of the CN-based ECs as determined by nitrogen physisorption techniques.

Table 4. The chemical composition of the surface of CN-based ECs (at.%) as determined by XPS.

Table 5. SSA of the CN-based ECs in the alkaline environment.